\documentclass[aps,prb,twocolumn,floats,showpacs]{revtex4}

\usepackage{bm}
\usepackage{graphicx}

\def \WHOZA{Institute of Experimental Physics, Faculty of Physics, University of Warsaw, Ho\.za 69, 00-681 Warsaw, Poland}

\begin{document}

\title{Efficient injection of spin-polarized excitons and optical spin orientation of a single Mn$^{2+}$ ion in a CdSe/ZnSe quantum dot}
\author{T. \surname{Smole\'nski}}\email{Tomasz.Smolenski@fuw.edu.pl}\affiliation{\WHOZA}
\author{W. \surname{Pacuski}}\affiliation{\WHOZA}
\author{M. \surname{Goryca}}\affiliation{\WHOZA}
\author{M. \surname{Nawrocki}}\affiliation{\WHOZA}
\author{A. \surname{Golnik}}\affiliation{\WHOZA}
\author{P. \surname{Kossacki}}\affiliation{\WHOZA}

\date{\today}

\begin{abstract}
Circularly polarized optical excitation is used to demonstrate the efficient injection of spin-polarized excitons to individual self-assembled CdSe quantum dots in ZnSe barrier. The exciton spin-transfer is studied by means of polarization-resolved single dot spectroscopy performed in magnetic field applied in Faraday configuration. Detailed analysis of the neutral exciton photoluminescence spectra reveals the presence of exciton spin relaxation during its lifetime in a quantum dot. This process is seen for both nonmagnetic dots and those containing single Mn$^{2+}$ ions. Taking this into account we determine the spin-polarization degree of excitons injected to a dot under circularly polarized below-the-barrier optical excitation at 488~nm. It is found to be close to 40\% in the entire range of the applied magnetic field. Exploiting the established spin-conserving excitation channel we demonstrate the optical spin orientation of a single Mn$^{2+}$ ion embedded in a CdSe/ZnSe quantum dot.
\end{abstract}

\pacs{78.67.Hc, 78.55.Et, 75.75.-c}

\maketitle

\section{Introduction}
Studies of single objects in a solid have recently attracted a lot of research attention. This led to coining the term \textit{solotronics} -- optoelectronics based on solitary dopants\cite{Flatte_NM_2011,Awschalom_S_2013,Rossier_NM_2013,Kobak_2014}. In order to achieve the control over a single dopant, in particular over its spin, different techniques were employed, depending on the studied system. For example, a single phosphorus center on Si surface could be controlled electrically, allowing for realization of a single atom transistor\cite{Fuechsle_NN_2012} and qubit\cite{Pla_N_2012, Buch_NC_2013}. Control of a single N-V center in diamond\cite{Jelezko_PRL_2004} can be obtained using a combination of internal optical and microwave transitions, which together enabled development of a room-temperature qubit\cite{Balasubramanian_N_09,Dolde_NP_13} and can be used for nanosensing of local magnetic fields\cite{Cooper_NC_2014}. Individual lanthanide ions such as Ce, Er or Pr can be addressed using intra-shell 4$f$, or inter-shell 4$f$-5$d$ optical transitions\cite{Yin_N_2013,Siyushev_NC_2014,Utikal_NC_2014}.
 
Transition metal ions with partially filled $d$-shell embedded in semiconductor offer another attractive possibility of optical manipulation mediated by spin-polarized carriers, which interact with the magnetic ion due to the $s$,$p$-$d$ exchange interaction\cite{Gaj_book_2010}. Such interaction leads to particularly interesting effects when individual magnetic ion is embedded in a quantum dot (QD)\cite{Besombes_PRL_2004}. In such case, the excitonic states are split due to the interaction with individual magnetic ion, allowing unambiguous readout of the magnetic ion spin state from polarization and energy of photon emitted from such system. So far four systems of QDs with single magnetic dopants have been reported: individual manganese ion in a CdTe\cite{Besombes_PRL_2004}, InAs\cite{Kudelski_PRL_2007}, and CdSe QD\cite{Kobak_2014, Smolenski_2014_arXiv}, and individual cobalt ion in a CdTe QD\cite{Kobak_2014}. The optical manipulation of the ion spin has been demonstrated only in the case of CdTe and InAs QDs containing manganese ions. It has been achieved either by strictly-resonant excitation of exchange-split excitonic states\cite{LeGall_PRB_2010, Baudin_PRL_2011} or through injection of spin-polarized excitons to Mn-doped QD\cite{LeGall_PRL_2009, Goryca_PRL_2009, Goryca_2010_Phys_E}. Such injection typically requires efficient spin-conserving excitation channels. They have been found for CdTe/ZnTe QDs with the use of resonant excitation of either an excited excitonic state in Mn-doped QD\cite{LeGall_PRL_2009}, or a ground state of neutral exciton in an adjacent, spontaneously coupled nonmagnetic dot\cite{Goryca_PRL_2009,tkaz_prb_2009,Goryca_2010_Phys_E,Koperski_PRB_2014}. However, the possibility of such resonant excitation has not been demonstrated for a CdSe/ZnSe QD containing a single manganese ion -- a new promising solotronic system with very long Mn$^{2+}$ spin relaxation time \cite{Kobak_2014}. Nevertheless, self-organized CdSe/ZnSe QDs offer a possibility to employ non-resonant, but still spin-conserving optical excitation channels, as it was shown for QD ensembles\cite{Kusrayev_PRB_2005}. 

In this work we show that spin-polarized excitons can be efficiently injected to individual self-assembled CdSe/ZnSe quantum dots by below-the-barrier optical excitation at 488~nm. Utilizing this technique we demonstrate the optical orientation of a single Mn$^{2+}$ ion spin in a CdSe/ZnSe QD. The paper is organized as follows. First, we present the sample growth procedure and describe the experimental setup in section \ref{sample_setup}. Section \ref{spectrum} contains a discussion of a photoluminescence spectrum of an individual CdSe/ZnSe QD. Next, we demonstrate the efficient optically-induced exciton spin transfer for CdSe/ZnSe QDs, both nonmagnetic (Sec. \ref{transfer}) and containing single Mn$^{2+}$ ions (Sec. \ref{transfer_Mn}). Finally, the optical orientation of Mn$^{2+}$ ion spin in a CdSe/ZnSe QD is presented and discussed in Sec. \ref{Mn_orientation}.

\section{Sample and experimental setup}
\label{sample_setup}

\begin{figure}
\includegraphics{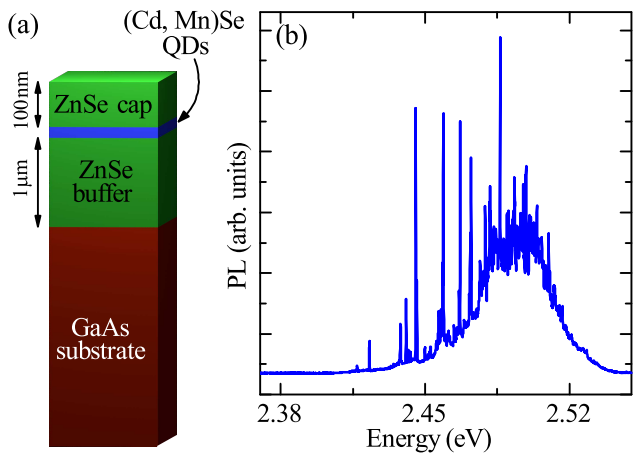}
\caption{(Color online)
(a) Cross-section scheme of the studied sample containing CdSe/ZnSe QDs with Mn$^{2+}$ ions (not in scale). (b) Micro-photoluminescence spectrum of CdSe/ZnSe QDs measured at randomly selected position on the sample surface, at helium temperature.
 \label{fig1}}
\end{figure}

The sample studied in this work contains a single layer of self-assembled CdSe/ZnSe QDs doped with manganese ions. It is grown using molecular beam epitaxy (MBE) on GaAs (100) substrate. The growth procedure is particularly simple. First, we grow 1~$\mu$m ZnSe buffer layer, then 2~monolayers of (Cd,Mn)Se which are immediately transformed into QDs, and finally 100~nm of ZnSe cap. Scheme of the sample cross-section is shown in Fig. \ref{fig1}(a). Formation of QDs is evidenced by photoluminescence (PL) measurements presented in this work. In particular, sharp QD emission lines are well resolved in a broad energy range micro-photoluminescence spectrum shown in Fig. \ref{fig1}(b).

In contrast to the growth of the first CdSe QDs with single manganese ions,\cite{Kobak_2014} in this work the substrate is kept at constant temperature of 300$^\circ$~C during the growth, and neither CdSe deposition by atomic layer epitaxy nor the amorphous selenium method\cite{Tinjod_2003_APL} is used. Molecular flux of manganese is optimized to assure high probability of finding QDs with exactly one Mn$^{2+}$ ion in each dot. 

Our experiments are carried out in a high-resolution micro-photoluminescence setup described in Refs. \onlinecite{tkaz_prb_2010} and \onlinecite{tkaz_prb_2011}. The sample is immersed in a superfluid helium inside a magneto-optical cryostat at temperature of about 1.8 K. The cryostat is equipped with a split-coil superconducting magnet producing magnetic field in the range of 0-10~T. The QDs are excited quasi-resonantly by a continues-wave Argon ion laser (lines $\lambda = 476.5$~nm or $\lambda = 488.0$~nm). A reflection microscope objective, immersed in a superfluid helium together with the sample, provides a spatial limitation of the PL excitation and detection to an area of diameter smaller than 1 $\mu$m. The QDs PL is resolved using a 0.5 m monochromator and recorded by a CCD camera.

In this study we investigate the properties of single CdSe/ZnSe QDs which do not contain magnetic ions and single CdSe/ZnSe QDs with single magnetic ions, both found in the same sample. Due to the high density of typical self-organized QDs, even for relatively small laser spot size a large number of dots are optically excited. This results in an inhomogeneously broadened QD ensemble PL spectrum with a characteristic line structure, which is shown in Fig. \ref{fig1}(b). In majority of the previous studies on selenide dots, an observation of isolated QD PL typically required additional sample processing, such as fabrication of masks or mesas \cite{Kulakovskii_PRL_1999, Patton_PRB_2003, Koudinov_PRB_2004, Akimov_PRB_2005}, which might affect the optical properties of selected QD (e.g., the polarization of emitted light). In contrast, our unstructured sample exhibits significantly lower QD density (which is apparent from the relatively low number of emission lines present in wide PL spectrum in Fig. \ref{fig1}(b)). In particular, it enables us to easily distinguish well-resolved emission lines of individual QDs in a long-wavelength tail of the PL band.

\section{PL spectrum of a single CdSe/ZnSe QD}
\label{spectrum}

\begin{figure}
\includegraphics{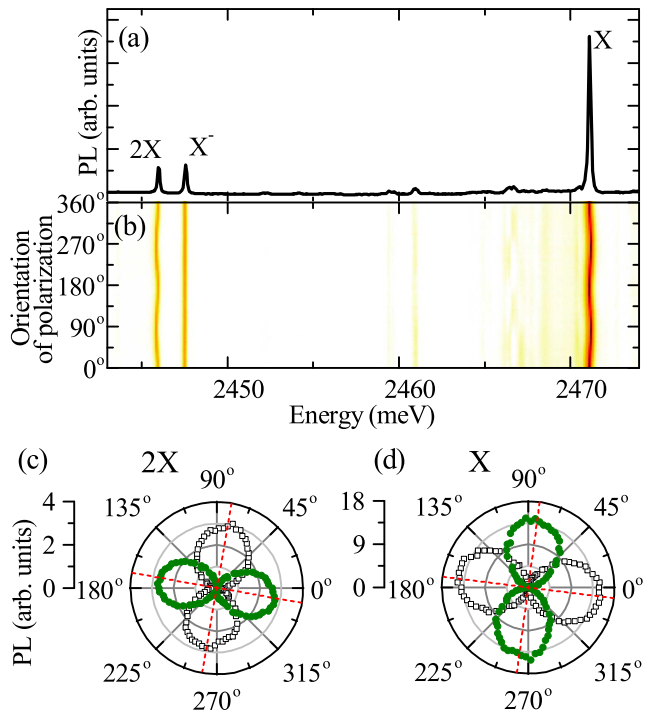}
\caption{(Color online)
(a) A PL spectrum of a single CdSe/ZnSe QD. (b) False-color plot presenting PL spectra of the same QD measured for different directions of detected linear polarization. (c), (d) Polar plots presenting anisotropy-related dependence of intensity of 2X and X emission lines doublets on the orientation of detected linear polarization for the same QD. Full symbols are related to the higher-energy lines of each doublet, while open ones - to the lower-energy lines. Dashed lines mark the orientations of the linear polarization of each transition.
 \label{fig2}}
\end{figure}

An example PL spectrum of an individual, nonmagnetic CdSe/ZnSe QD is shown in Fig. \ref{fig2}(a). It consists of three sharp emission lines, which are identified as originating from recombination of neutral exciton (X), negatively charged exciton (X$^-$) and biexciton (2X). The identification is based on the in-plane optical anisotropy and relative emission energies. The sequence and relative energies of observed excitonic transitions are found to be similar to ones previously reported for self-assembled CdSe/ZnSe QDs \cite{Patton_PRB_2003, Kobak_2014}. On this basis the highest and lowest energy emission lines are first tentatively identified as related to the recombination of neutral exciton and biexciton, respectively. This attribution is further confirmed by a characteristic fine structure of both invoked transitions, which is a result of the in-plane anisotropy of a QD confining potential (typical for self-organized dots \cite{Gammon_PRL_1996, Bayer_PRB_2002, tkaz_prb_2011}). More specifically, the anisotropic part of exchange interaction between an electron and a heavy hole splits two bright states of neutral exciton by $\delta_1$ energy leading to the presence of two orthogonally linearly polarized X emission lines. Since X is a final state of a spin-singlet biexciton recombination, the 2X optical transitions also form a linearly polarized doublet split by the same energy $\delta_1$ as X emission lines. These predictions are reproduced by the experimental data in Fig. \ref{fig2}(b), which shows PL spectra of the analyzed QD measured for different orientations of detected linear polarization. Due to relatively small $\delta_1\simeq0.10$~meV anisotropy splitting compared to X (or 2X) emission line FWHM $\simeq0.19$~meV, the described fine structure manifest itself only as a small oscillations of the central (average) energy of X and 2X optical transitions. Nevertheless, fitting emission line with two Gaussian profiles of fixed energies enabled us to determine the intensities of each component of X and 2X transitions. The angular dependences of the obtained intensities are shown in Figs. \ref{fig2}(c), \ref{fig2}(d). As expected, in both cases the fine-split emission lines are fully linearly polarized in perpendicular directions, which are similar for X and 2X emission lines.

The third, middle-energy emission line present in the QD PL spectrum does not exhibit any noticeable fine structure (Fig. \ref{fig2}(b)). This is typical for recombination of a trion,\cite{Koudinov_PRB_2004, Leger_PRB_2007, tkaz_prb_2011} where both the initial state (with two majority carriers forming a closed shell) and the final state (consisting of one residual carrier) are strictly degenerate. Basing on this observation as well as on the comparison of relative energies of QD emission lines with the results of previous studies,\cite{Patton_PRB_2003, Kobak_2014} we attribute the middle-energy line to a singly charged exciton. The identification of its negative sign (as marked in Fig. \ref{fig2}(a)) is first tentatively based on a typical tendency for selenide materials to be n-type. The sign attribution is unequivocally confirmed by an observation of negative optical polarization transfer for X$^{-}$ transition under quasi-resonant excitation, which is discussed in section \ref{transfer:tion} of this paper.

\section{Spin-transfer for a CdSe/ZnSe QD}
\label{transfer}

In this section we focus on the ability to control over the spin of excitons injected to a single CdSe/ZnSe QD. In particular, we demonstrate that optical excitation of a CdSe QD with circularly polarized light of photon energy appropriately close to the QD emission energy provides a significant spin-polarization of excitons captured by the dot. In order to investigate the efficiency of spin-transfer we performed polarization-resolved measurements of QD PL under below-the-barrier CW excitation at 476.5~nm or 488~nm in external magnetic field (Faraday configuration). We studied several, randomly selected dots. Here we present data obtained for the representative dot discussed in the previous section.

\subsection{Neutral exciton}

\begin{figure}
\includegraphics{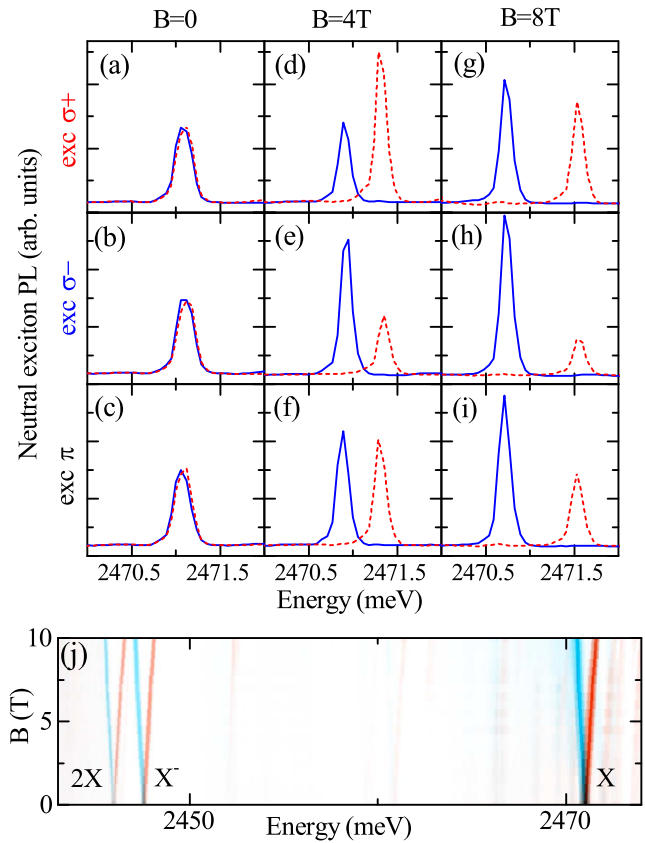}
\caption{(Color online) (a)-(i) Neutral exciton PL spectra detected in $\sigma^+$ (dashed red lines) and $\sigma^-$ (solid blue lines) circular polarizations under quasi-resonant excitation at 488~nm with indicated circular or linear polarizations. The measurements were performed at magnetic field $B=0$ (a)-(c), $B=4$~T (d)-(f) and $B=8$~T (g)-(i) applied along the QD growth axis. (j) The QD PL spectrum as a function of the magnetic field in Faraday configuration. Color saturation denotes degree of circular polarization (red -- $\sigma^+$, blue -- $\sigma^-$), while brightness indicates the PL intensity.
 \label{fig3}}
\end{figure}

We first analyze the PL of the neutral exciton. Figures \ref{fig3}(a)-\ref{fig3}(i) show the X PL spectra at $B=0$, 4~T and 8~T measured in two circular polarizations of detection under circularly or linearly polarized excitation at 488~nm ($\simeq 2.55$~eV). In the absence of the magnetic field no circular polarization of X PL is observed, independently of the polarization of excitation (Figs. \ref{fig3}(a)-\ref{fig3}(c)). This is related to the fine-structure splitting of the neutral exciton state. Even if under circular polarization of excitation a spin-polarized exciton is injected to the QD, its initial state is a coherent superposition of two $\delta_1$-split linearly polarized X eigenstates in the QD. Therefore, the exciton is undergoing oscillations between its eigenstates. The period of these oscillations $2\pi\hbar/\delta_1\simeq40$~ps is several times shorter than the neutral exciton lifetime, which is of the order of a few hundreds of picoseconds in the case of CdSe/ZnSe QDs\cite{Patton_PRB_2003, Kobak_2014, Smolenski_2014_arXiv}. As a consequence, the time-integrated X PL is almost completely unpolarized, as observed in the experiment.

The situation is qualitatively different after application of the magnetic field along the QD growth axis. The field introduces a Zeeman splitting between two spin-polarized states of the neutral exciton. At the sufficiently large field this splitting of $g\mu_BB$ (where $g$ is the excitonic g-factor) dominates over the X fine-structure splitting $\delta_1$ and exciton eigenstates in the QD become nearly pure spin states. Experimentally, it manifest itself in the presence of almost fully circularly polarized X emission lines. In the case of the studied QD, it is visible even in the field of a few Tesla (Fig. \ref{fig3}(j)) due to relatively small $\delta_1$ splitting. Thus, the application of such field enables us to determine the spin-polarization degree of excitons injected to the QD from the X PL spectra. Such spectra measured at $B=4$~T are shown in Figs. \ref{fig3}(d)-\ref{fig3}(f). The X PL intensities obtained for the same circular polarization of excitation and detection (co-polarized) are clearly larger than those measured in different circular polarizations (cross-polarized). This observation unequivocally confirms the presence of an efficient optically-induced exciton spin transfer, both for $\sigma^+$ and $\sigma^-$ polarized excitation at 488~nm.

Under linearly polarized excitation at $B=4$~T we also observe a small, but not negligible degree of X PL circular polarization (Fig. \ref{fig3}(f)). Independently of the orientation of linear polarization of excitation, the lower-energy X emission line ($\sigma^-$ polarized) has slightly larger intensity compared to the higher-energy one ($\sigma^+$ polarized). This effect is more pronounced at higher magnetic field of 8~T (Fig. \ref{fig3}(i)). In order to explain its origin, we first notice that excitation with linear polarization is unlikely to induce any spin-polarization of excitons injected to the QD (unimportance of, in principle possible, linear-to-circular polarization conversion \cite{Astakhov_PRL_2006, Kusrayev_PRB_2005, tkaz_prb_2009} will be argued in the latter part of this section). Thus, we relate the observed effect to an exciton spin-flip process most probably assisted by an acoustic phonon. As the phonon emission is strongly favored over the phonon absorption at low temperatures, the spin-flip causes exciton relaxation towards the lower-energy state during X lifetime in the QD. As a consequence, $\sigma^-$ polarized emission line of the neutral exciton has enhanced intensity even under excitation providing no spin-polarization of optically-created excitons. Since the spin-flip rate exhibits a strong dependence on the energy of emitted phonon, the efficiency of X relaxation increases with an increasing Zeeman splitting. It underlies the mentioned above increase of the intensity ratio of lower- and higher-energy X emission lines at higher magnetic field under linearly polarized excitation. The influence of exciton spin-flip is also visible in the X PL spectra measured at $B=8$~T with circular polarization of excitation (Figs. \ref{fig3}(g), \ref{fig3}(h)). In this case, the intensities of X emission lines are governed by an interplay of the optically-induced exciton spin-transfer and phonon-assisted X spin relaxation. In particular, the excitation with $\sigma^-$ ($\sigma^+$) polarization results in an increased (decreased) intensity ratio of $\sigma^-$ and $\sigma^+$ polarized X emission lines compared to the ratio obtained under linearly polarized excitation.

\begin{figure}
\includegraphics{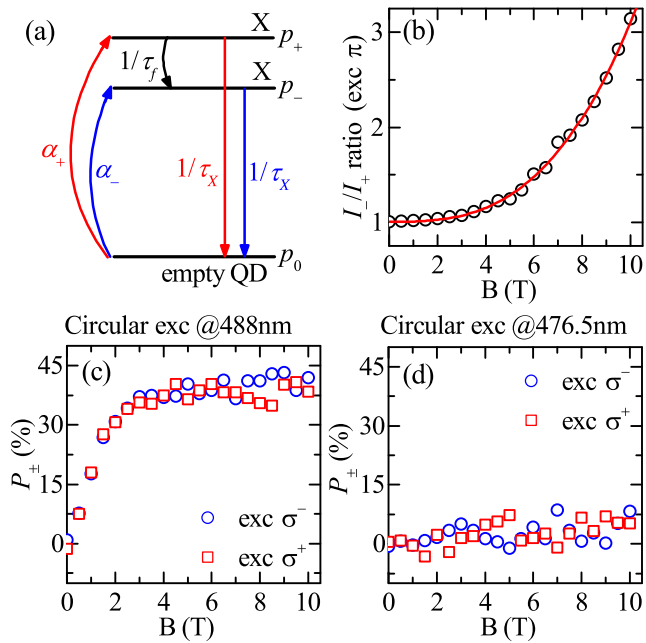}
\caption{(Color online) (a) A schematic illustration of QD states and transitions considered in the rate-equation model. (b) Ratio of lower- and higher-energy X emission line intensities under linearly polarized excitation as a function of the magnetic field. The solid line represents the fitted curve accounting for power dependence of the X spin-flip rate on the energy splitting of X states. (c), (d) Degrees of spin-transfer efficiency described by Eq. (\ref{eq:transfer_degree}) determined for excitation at 488~nm, 476.5~nm with $\sigma^+$ (squares) and $\sigma^-$ (circles) polarization.
\label{fig4}}
\end{figure}

In order to perform a quantitative analysis of the exciton spin-transfer efficiency at different values of the magnetic field, we introduce a simple rate-equation model. It includes the probabilities of occupation of three states: the empty QD ($p_0$) and the neutral exciton in lower- or higher-energy state ($p_{-}$ or $p_{+}$, respectively). We do not take into account the biexciton state, as the excitation power used in our experiment is sufficiently low to assure that 2X emission intensity is not exceeding a few percent of the X intensity. In general, we expect that the injection of excitons to the dot is a complex process which may involve capture of both single carriers and electron-hole pairs \cite{Suffczynski_PRB_2006, tkaz_prb_2010}. Since the detailed analysis of this process remains beyond the scope of this paper, we describe it by introducing two effective rates $\alpha_+$ and $\alpha_-$, which account for injection of an exciton to the higher- and lower-energy state, respectively. Moreover, we take into account the exciton radiative lifetime $\tau_X$, which is assumed to be the same for both considered X states and independent of the magnetic field. Finally, we include the spin-flip process turning a higher-energy exciton into a lower-energy one by introducing its characteristic time $\tau_f$, which depends on the energy splitting of both invoked states. The scheme of the states and transitions is shown in Fig. \ref{fig4}(a).
Within our model, the steady state is given by the following linear equations
\begin{eqnarray}
\frac{dp_+}{dt}&=&-\frac{p_+}{\tau_X}-\frac{p_+}{\tau_f}+\alpha_+p_0=0,\\
\frac{dp_-}{dt}&=&-\frac{p_-}{\tau_X}+\frac{p_+}{\tau_f}+\alpha_-p_0=0,\\
\frac{dp_0}{dt}&=&\frac{p_++p_-}{\tau_X}-(\alpha_++\alpha_-)p_0=0.
\end{eqnarray}
As $p_-$ and $p_+$ probabilities are proportional to the intensities $I_-$ and $I_+$ of lower- and higher-energy X emission lines, their ratio $I_-/I_+$ can be easily determined from the above set of equations and reads
\begin{equation}\label{eq:intensity_ratio}
\frac{I_-}{I_+}=\left(1+\frac{\alpha_-}{\alpha_+}\right)\left(1+\frac{\tau_X}{\tau_f}\right)-1.
\end{equation}

In the case of linearly polarized excitation, the probabilities $\alpha_+$ and $\alpha_-$ of an exciton injection to both considered X states in the QD are equal. Thus the ratio of X intensities under such an excitation is given by $I_-/I_+=1+2\tau_X/\tau_f$, allowing a direct determination of relative neutral exciton spin-flip rate. Fig. \ref{fig4}(b) presents the measured $I_-/I_+$ ratio as a function of the magnetic field. In accordance to the above considerations we observe a clear, superlinear increase of the relative intensity of lower-energy X emission line resulting from the shortening of X spin-flip time at higher fields. Detailed analysis of the obtained results reveals the power dependence of the spin-flip rate $1/\tau_f$ on the energy splitting $\Delta E(B)$ of neutral exciton states in the dot. In particular, the curve given by $I_-/I_+=1+a\Delta E^\gamma$ perfectly reproduces the experimental data from Fig. \ref{fig4}(b) for $\gamma=2.95\pm0.05$. This demonstrates that within our experimental precision, the X spin-flip rate $1/\tau_f$ is proportional to the third power of X energy splitting, as previously theoretically predicted for a single phonon-mediated X spin relaxation in a QD \cite{Tsitsishvili_PRB_2003}.

Taking into account the determined X spin-flip rate dependence on the magnetic field we can extract the exciton spin-transfer efficiency under circularly polarized excitation. A quantitative measure of this efficiency is a degree to which the excitons are injected to one of X eigenstates in the QD. Using Eq. (\ref{eq:intensity_ratio}) it can be expressed as
\begin{equation}\label{eq:transfer_degree}
P_\pm=\frac{\alpha_\pm-\alpha_\mp}{\alpha_++\alpha_-}=\pm\left(2\frac{1+\frac{\tau_X}{\tau_f}}{1+\frac{I_-}{I_+}}-1\right),
\end{equation}
where $P_+$ ($P_-$) corresponds to the excitation with $\sigma^+$ ($\sigma^-$) circular polarization. It should be noted that at sufficiently large magnetic field (i.e., when nearly pure spin states are eigenstates of X) the above-introduced $P_\pm$ accounts directly for spin-polarization degree of excitons injected to the dot. Moreover, for such fields $P_\pm$ would be equal to the degree of X PL circular polarization in the case of an absence of X spin-flip. On the contrary, at lower magnetic fields ($B\lesssim2$~T for the studied QD) the interpretation of $P_\pm$ is not as straightforward since X eigenstates are partially linearly polarized.

The evolution of the typical measured $P_\pm$ degree in the magnetic field is shown in Fig. \ref{fig4}(c). The exciton spin-transfer efficiency is found to be similar for both circular polarizations of excitation. At $B>3$~T we observe high spin-polarization degree of injected excitons, which is almost constant and exceeds 40\%. This result unequivocally confirms the pronounced conservation of exciton spin-polarization under excitation of the studied CdSe/ZnSe QD at 488~nm. Similar effect is observed for various quantum dots independently of the excitonic emission energy. On the other hand, at $B<3$~T the $P_\pm$ degree drops due to raising influence of the anisotropy-related excitonic fine structure, which prevents precise determination of the spin-transfer efficiency at lower fields. Interestingly, the reduction of excitation wavelength from 488~nm to 476.5~nm ($\simeq 2.61$~eV) results in an almost complete loss of the exciton spin-transfer in the entire range of the magnetic field (Fig. \ref{fig4}(d)).

It is noteworthy that our findings do not indicate the presence of any effective linear-to-circular polarization conversion, which was observed in the previous studies on an ensemble of CdSe/ZnSe QDs\cite{Kusrayev_PRB_2005}. Such a conversion would lead to a non-zero spin-polarization degree of excitons injected to a QD under linearly polarized excitation. In such a scenario, the intensities of X emission lines under this excitation would not be solely related to the X spin-flip process. As a consequence, its rate would be imprecisely determined, which would finally result in a difference between $P_+$ and $P_-$ degrees obtained for both circular polarizations of excitation from Eq. (\ref{eq:transfer_degree}). However, almost equal values of both degrees determined experimentally reveals unimportance of the linear-to-circular polarization conversion in case of the studied QDs.

\subsection{Trion and biexciton}
\label{transfer:tion}

\begin{figure}
\includegraphics{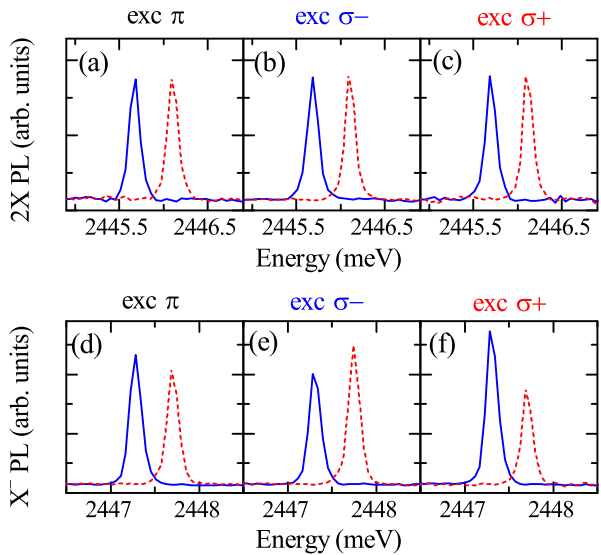}
\caption{(Color online) The PL spectra of biexciton (a)-(c) and negatively charged trion (d)-(f) measured at $B=4$~T under excitation at 488~nm with indicated polarizations. The spectra were detected in $\sigma^-$ (solid blue lines) and $\sigma^+$ (dashed red lines) circular polarizations.
\label{fig5}}
\end{figure}

In order to provide a comprehensive description of polarization properties of CdSe/ZnSe QD PL under optical excitation at 488~nm, we briefly analyze the spectra of two remaining excitonic complexes: biexciton and negatively charged trion. In particular, we focus on the measurements carried out at magnetic field of 4~T (Faraday configuration). The corresponding 2X and X$^{-}$ PL spectra obtained with polarization-resolution under circularly or linearly polarized excitation are shown in Fig. \ref{fig5}. 

In case of the biexciton (Figs. \ref{fig5}(a)-\ref{fig5}(c)), the PL intensities are not affected by switching the polarization of excitation. It is related to the fact that 2X consists of two electron-hole pairs forming closed shells. As such, it cannot carry any spin information and recombines to one out of two Zeeman-split neutral exciton states with the same probabilities. As a result, we observe equal intensities of both 2X emission lines independently of the polarization of excitation.

The physical picture in case of the trion is more complex (Figs. \ref{fig5}(d)-\ref{fig5}(f)). Under linearly polarized excitation providing no spin-transfer of injected excitons, the lower-energy emission line is slightly more intense (Fig. \ref{fig5}(d)). Similarly to the case of the neutral exciton, this effect is interpreted as originating from spin relaxation taking place during the trion lifetime in the QD. However, such relaxation has qualitatively different character compared to the neutral exciton spin-flip process analyzed in the previous section. In particular, it is directly related to a spin-flip of a single hole between its Zeeman-split states, as the remaining two electrons forming negatively charged trion are accommodated in a closed shell and cannot participate in the spin relaxation. The presence of the hole spin-flip affects also X$^-$ PL spectra measured under circularly polarized excitation increasing the intensity of the lower-energy line (Figs. \ref{fig5}(e), \ref{fig5}(f)). Despite this effect, in such case a pronounced optically-induced spin-transfer is observed. But, in contrast to the neutral exciton, the spin transfer has negative sign since the cross-polarized X$^-$ emission is more intense compared to the co-polarized emission. The apparent negative optical orientation for X$^-$ was previously observed in different QD systems, both III-V \cite{Cortez_PRL_2002, Laurent_PRB_2006} and II-VI \cite{Akimov_PRL_2006, tkaz_prb_2009}. Following Ref. \onlinecite{Cortez_PRL_2002}, we interpret it as arising due to possible spin mismatch between an electron residing in the dot and the one forming the spin-polarized injected exciton. If both electrons have the same spin, the Pauli blockade prohibits complete energy relaxation of the injected exciton. In such a case, spin-conserving relaxation processes stop at the level of hot trion with one electron in the ground state and the other occupying the first excited state. The hot trion state is typically characterized by relatively strong anisotropic part of the electron-hole exchange interaction \cite{Akimov_PRB_2005, tkaz_prb_2011, tkaz_prb_2013}, which accompanied by the Fr\"{o}hlich interaction between electron pair and LO phonon \cite{Benny_PRB_2014} leads to simultaneous spin-flip of the electron and the hole (flip-flop). As a result, the spin of injected exciton is effectively reversed, which finally underlies the negative optical spin-transfer determined experimentally for X$^-$. On the other hand, no such effect has been observed for positively charged trion due to much lower efficiency of the electron-hole flip-flop process in this case \cite{Benny_PRB_2014}. Thus, our findings confirm the negative charge state of the only trion which emission line is visible in the PL spectra of the studied QDs.

\section{Spin-transfer for Mn-doped CdSe/ZnSe QD}
\label{transfer_Mn}

\begin{figure}
\includegraphics{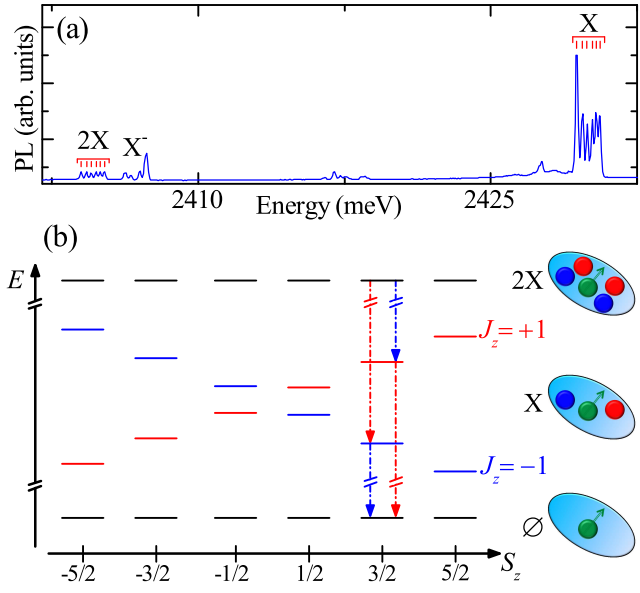}
\caption{(Color online) (a) A PL spectrum of a CdSe/ZnSe QD containing a single Mn$^{2+}$ ion. (b) Simplified diagram of neutral exciton and biexciton energy levels in Mn-doped QD at zero magnetic field. The states are displayed as a function of their energy and Mn$^{2+}$ ion spin projection on the QD growth axis. Color of the neutral exciton levels denotes the X spin projection on the growth axis, as well as the circular polarization of corresponding X optical transitions (red -- $\sigma^+$, blue -- $\sigma^-$). For clarity, only the transitions related to 3/2 spin projection of the Mn$^{2+}$ ion are shown. 
\label{fig6}}
\end{figure}

The efficient injection of spin-polarized excitons is also demonstrated for CdSe/ZnSe QDs containing single Mn$^{2+}$ ions. The typical PL spectrum of an example Mn-doped dot is shown in Fig. \ref{fig6}(a). Similarly to the case of a nonmagnetic QD (see Fig. \ref{fig2}(a)), we observe emission lines originating from the recombination of X, X$^{-}$ and 2X. However, each of them exhibits multifold splitting related to the presence of $s,p$-$d$ exchange interaction between confined carriers and Mn$^{2+}$ ion (total spin $S=5/2$) \cite{Besombes_PRL_2004, Leger_PRB_2007, Leger_PRL_2006, Goryca_PRB_2010, Kobak_2014}. Due to heavy hole character of the hole ground state in a QD, the hole-Mn exchange is effectively Ising-like \cite{Besombes_PRL_2004}. As a result, the neutral exciton emission line is split into six components as observed in Fig. \ref{fig6}(a). Each of them is twofold degenerated and related to a particular spin projection of the X-Mn system on the QD growth axis $z$ (which is the quantization axis of the hole). More specifically, the consecutive energy levels of the neutral exciton with spin projection $J_z=\pm1$ (coupled to $\sigma^\pm$ polarized light) correspond to subsequent Mn$^{2+}$ spin projections $S_z$ equal to $\mp5/2$, $\mp3/2$, $\mp1/2$, $\pm1/2$, $\pm3/2$, and $\pm5/2$ (as schematically depicted in Fig. \ref{fig6}(b)). The same underlying exchange splitting of the neutral exciton affects also the biexciton emission spectrum, since this excitonic complex is a spin-singlet and does not interact with the ion. As a consequence, the 2X state is degenerate (Fig. \ref{fig6}(b)) and its emission line is split only due to the final state of the recombination (i.e., the neutral exciton state). Finally, the biexciton PL exhibits similar sixfold splitting as the neutral exciton, which is clearly visible in Fig. \ref{fig6}(a).

\begin{figure}
\includegraphics{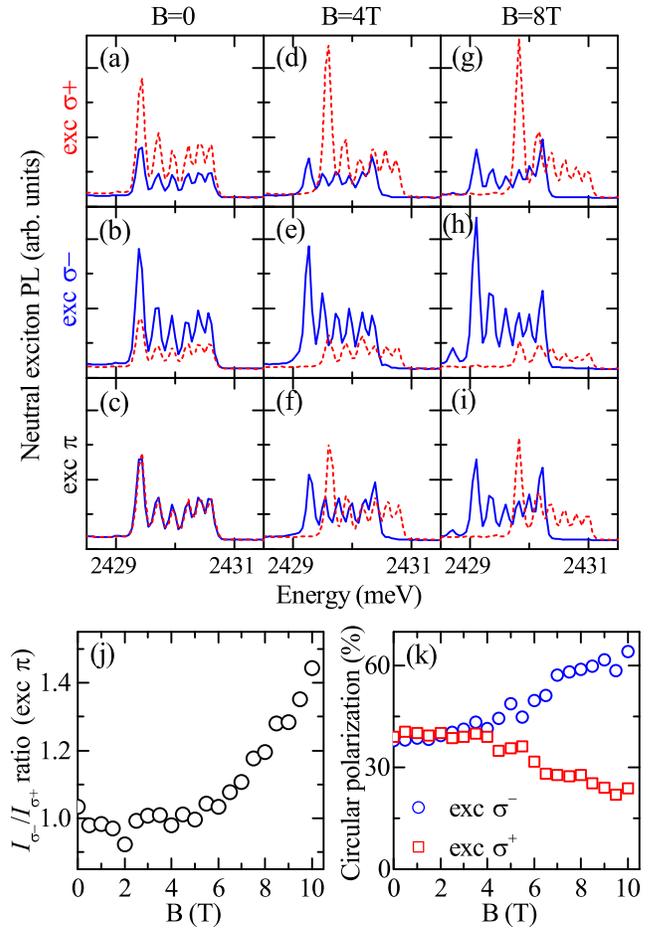}
\caption{(Color online) (a)-(i) PL spectra of the neutral exciton confined in Mn-doped QD under below-the-barrier excitation at 488~nm with indicated circular or linear polarizations. The spectra were detected in $\sigma^+$ (dashed red lines) and $\sigma^-$ (solid blue lines) circular polarizations and measured at magnetic field $B=0$ (a)-(c), $B=4$~T (d)-(f) and $B=8$~T (g)-(i). (j) Magnetic field dependence of the ratio of the integrated X PL intensities detected in $\sigma^-$ and $\sigma^+$ circular polarizations under linearly polarized excitation. (k) Circular polarization degree of the integrated X PL under excitation at 488~nm with $\sigma^+$ (squares) and $\sigma^-$ (circles) polarization as a function of the magnetic field.
\label{fig7}}
\end{figure}

The efficiency of exciton spin-transfer under excitation at 488~nm of the Mn-doped QD is determined in polarization-resolved measurements of the X PL in external magnetic field applied in Faraday geometry. The results are presented in Figs. \ref{fig7}(a)-\ref{fig7}(i). In contrast to the case of nonmagnetic QD, a pronounced X spin-transfer is observed under circularly polarized excitation even in the absence of the magnetic field, as the co-polarized integrated intensities of X PL are clearly larger than the cross-polarized intensities (Figs. \ref{fig7}(a), \ref{fig7}(b)). This is related to a negligible influence of the anisotropy-related fine-structure on the neutral exciton states, split by the exchange interaction with the Mn$^{2+}$ ion, since $\delta_1$ is typically much smaller compared to X-Mn exchange-induced splitting\cite{Leger_PRL_2005}. As a result, even in the presence of the fine structure nearly pure spin states are the eigenstates of X in Mn-doped QD. The Zeeman effect introduced by the magnetic field along the growth axis additionally splits the spin states of X-Mn system. As a consequence, X spin-transfer is observed in the entire range of the applied field (Figs. \ref{fig7}(d)-\ref{fig7}(e), \ref{fig7}(g)-\ref{fig7}(h)). This result confirms that circularly polarized optical excitation at 488~nm enables injection of spin-polarized excitons to a CdSe QD independently of the value of magnetic field. 

The average value of X PL circular polarization degree under circularly polarized excitation is found to be approximately field-independent and equal to about 40\% for all studied Mn-doped QDs. However, this value is not necessarily identical to exciton spin-transfer efficiency. More specifically, the quantitative determination of this efficiency requires taking into account possible spin relaxation of X-Mn system occurring during the X lifetime in a QD. Such relaxation may concern both the exciton spin-flip between two states corresponding to the same projection of the ion spin as well as the processes involving simultaneous spin-flip of the X and the Mn$^{2+}$ ion. Their presence is revealed by zero-field X PL spectra measured under linearly polarized excitation (Fig. \ref{fig7}(c)). In such a case, no spin polarization of the Mn$^{2+}$ ion is expected, which should manifest itself in equal intensities of six emission lines corresponding to different projections of the ion spin. On the contrary, the intensity of the low-energy emission line is significantly larger compared to the higher-energy lines, independently of the circular polarization of detection. Since the low-energy line corresponds to $\pm5/2$ ion spin projection in $\sigma^\mp$ polarization of detection, this effect cannot be attributed to a steady state polarization of the ion spin. On the other hand, it may originate from X spin-flip process causing the exciton relaxation from $|J_z=\pm1,S_z=\pm5/2\rangle$ state towards $|J_z=\mp1,S_z=\pm5/2\rangle$. As this process does not affect the ion spin, its underlying mechanism is identical as the one entailing exciton spin-flip in a nonmagnetic QD between Zeeman-split states. The latter process is found to be quite efficient for a large energy splitting of the invoked states. Bearing in mind that energy difference between $|J_z=\pm1,S_z=\pm5/2\rangle$ and $|J_z=\mp1,S_z=\pm5/2\rangle$ states corresponds to energy splitting of two extreme X PL lines (equal to about 1.2~meV in the case of the studied QD), the proposed spin-relaxation can play a significant role in Mn-doped QD. In particular, it increases the intensity of the low-energy X emission line. On the other hand, its presence would also lead to a decrease of the high-energy line intensity. However, the experimental results (Fig. \ref{fig7}(c)) do not reflect this prediction. Thus, we expect that other X-Mn spin relaxation processes have to be important. They might be related to a mutual flip of exciton and ion spins or a spin-flip of a single carrier forming the neutral exciton resulting in a creation of a dark exciton \cite{Cywinski_PRB_2010}. The precise identification and detailed description of these relaxation processes remain beyond of the scope of this paper and require further studies.

The application of the magnetic field modifies the energies of the X-Mn states (presented in Fig. \ref{fig6}(b)) introducing both exciton and ion Zeeman splittings. As a consequence, the symmetry between energy levels corresponding to $J_z=\pm1$ excitons is no longer present and X-Mn spin relaxation entails a difference of X PL intensities measured in two circular polarizations of detection under linearly polarized excitation (Figs. \ref{fig7}(f), \ref{fig7}(i)). As the relative energy of $J_z=-1$ exciton is decreased, the spin-flip processes increase the intensity of the corresponding $\sigma^-$ polarized X PL. It is clearly visible in Fig. \ref{fig7}(j) presenting the ratio of integrated X PL intensities detected in $\sigma^-$ and $\sigma^+$ polarizations under linearly polarized excitation. The determined ratio is however lower compared to the nonmagnetic QD for the same magnetic field (see Fig. \ref{fig4}(b)). The possible explanation of this effect is related to a similar strength of the X-Mn exchange interaction and the excitonic Zeeman effect at high magnetic field. In particular, even at $B=10$~T the Zeeman effect is too weak to invert the zero-field order of the exchange-split excitonic states corresponding to $-5/2$ ion spin projection. As a result, some of $|J_z=+1, S_z\rangle$ states have lower energy than $|J_z=-1, S_z\rangle$ even at high field. It finally leads to a suppression of the overall spin relaxation between $J_z=\pm1$ excitons when compared to a nonmagnetic QD. Nevertheless, the influence of the spin-flip is also visible in the X PL spectra measured under circularly polarized excitation. In such case, the spin relaxation results in an increase (decrease) of the circular polarization degree of integrated X PL under excitation with $\sigma^-$ ($\sigma^+$) polarization at high magnetic field (Fig. \ref{fig7}(k)). Therefore, the obtained degrees do not exactly correspond to the spin-polarization degrees of excitons injected to the dot. In principle, the latter might be extracted from a detailed analysis of X PL spectra measured under different polarizations of excitation. However, a complicated spin relaxation of the X-Mn system prevents us to perform such an analysis in a similar way as it is carried out in the case of nonmagnetic QD (described in section \ref{transfer}). We only note that the average degree of circular polarization of X PL under $\sigma^-$ and $\sigma^+$ polarized excitation of about $40\%$ defines the lower limit of the exciton spin-transfer efficiency in the entire range of the applied magnetic field.

\section{Optical spin orientation of Mn$^{2+}$ ion in a CdSe/ZnSe QD}
\label{Mn_orientation}

Exploiting the ability to inject the spin-polarized excitons to Mn-doped CdSe/ZnSe QD we study their influence on the Mn$^{2+}$ ion spin state. Such an influence so far has been demonstrated for CdTe/ZnTe QDs, in which the single Mn$^{2+}$ ion spin could be oriented towards $\pm5/2$ state via its exchange interaction with $J_z=\mp1$ excitons \cite{LeGall_PRL_2009, Goryca_PRL_2009}. In order to examine the presence of this phenomenon in the case of CdSe QDs, we first need to establish a method enabling unambiguous optical readout of the ion spin state. Following the previous studies of CdTe QDs one may think about determining the ion spin basing on the intensities of six emission lines of the neutral exciton detected in a given circular polarization. However, their values provide only information about the probabilities of finding the X-Mn system in a particular state at the moment of exciton recombination. If no spin relaxation is occurring during the X lifetime in a QD, those probabilities are indeed equivalent to the probabilities for the Mn$^{2+}$ ion to have one of its six possible spin projections in the absence of the exciton. On the contrary, the efficient spin-flip processes of X-Mn system present in the studied CdSe QDs make this approach no longer valid. For this reason we monitor the ion spin basing on the intensities of the biexciton emission lines. Since the 2X is a spin-singlet, it is decoupled from the ion and acts as an almost ideal probe of its spin state in an empty dot. A small interaction between the 2X and the Mn$^{2+}$ ion related to a configuration mixing\cite{Trojnar_PRB_2013} might be neglected here. As a consequence, the intensities of the 2X emission lines reflect the probabilities of finding the Mn$^{2+}$ ion in a spin state defined by the line energy and polarization of detection.

\begin{figure}
\includegraphics{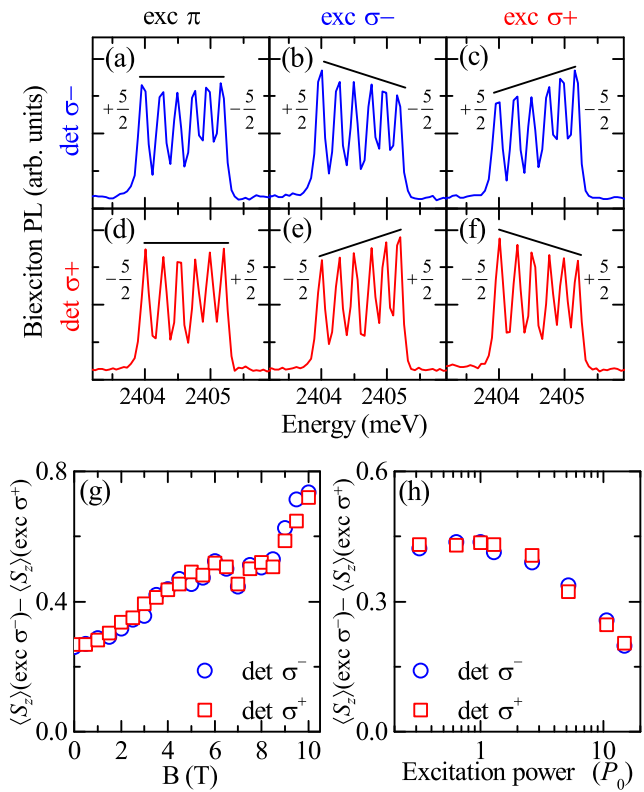}
\caption{(Color online) (a)-(f) The zero-field PL spectra of biexciton confined in Mn-doped QD. The spectra were excited at 488~nm with indicated circular and linear polarizations and detected in $\sigma^-$ (blue, (a)-(c)) and $\sigma^+$ (red, (d)-(f)) polarizations. The ion spin projections corresponding to the emission lines of extreme energies are indicated. Solid lines representing the direction of change of 2X intensities are drawn to guide the eye. (g),(h) The efficiency of optically-induced Mn$^{2+}$ spin orientation measured as a function of: magnetic field (g) and excitation power at $B=4$~T (h). Circles (squares) represent the efficiencies determined in $\sigma^-$ ($\sigma^+$) polarization of detection. The excitation power is expressed in $P_0$ units, where $P_0$ is a power used to obtain the data presented in (g). The saturation power of this QD corresponds to $30P_0$.
\label{fig8}}
\end{figure}

Having established the procedure of Mn$^{2+}$ spin readout in a CdSe QD, we verify the possibility of the ion spin manipulation through injection of spin-polarized excitons to the dot. To do so, we measure the biexciton PL spectra under circularly or linearly polarized excitation in a fixed circular polarization of detection. This ensures that each 2X emission line corresponds to a specified spin projection of the Mn$^{2+}$ ion. The zero-field spectra detected in $\sigma^-$ polarization are shown in Figs. \ref{fig8}(a)-\ref{fig8}(c). In the case of linearly polarized excitation, the intensities of all 2X emission lines are almost equal (Fig. \ref{fig8}(a)), which is a direct fingerprint of depolarized ion spin. On the other hand, excitation with circularly polarized light results in a small, but not negligible non-uniform distribution of the intensities between six lines. In particular, under $\sigma^-$ polarized excitation the lower-energy lines are more intense, while the higher-energy lines less (Fig. \ref{fig8}(b)). Switching the polarization of excitation to $\sigma^+$ leads to opposite effect, i.e., stronger higher-energy lines and weaker lower-energy lines (Fig. \ref{fig8}(c)). This observation unequivocally confirms the presence of optically-induced Mn$^{2+}$ ion spin polarization in a CdSe/ZnSe QD. Since $\sigma^-$ polarized 2X emission lines of subsequent energies correspond to Mn$^{2+}$ spin projections ranging from $5/2$ to $-5/2$, our findings indicate that injection of $J_z=\pm1$ excitons orients the ion spin towards $\mp5/2$ state, similarly to the case of Mn-doped CdTe QDs\cite{LeGall_PRL_2009, Goryca_PRL_2009, Goryca_2010_Phys_E}. The Mn$^{2+}$ spin can be also independently probed in $\sigma^+$ polarization of detection (Figs. \ref{fig8}(d)-\ref{fig8}(f)). In such case, the 2X emission energies correspond to subsequent ion spin projections in a reversed order, namely ranging from $-5/2$ to $5/2$. As the ion spin state depends only on the spin of injected excitons, we observe opposite change of 2X emission lines intensities under the same circular polarization of excitation compared to the previously used $\sigma^-$ polarization of detection, which is clearly seen in Figs. \ref{fig8}(e)-\ref{fig8}(f).

In order to perform a detailed analysis of the optical spin orientation efficiency, we introduce its quantitative measure defined as a difference between mean values of the ion spin projection obtained under $\sigma^-$ and $\sigma^+$ polarized excitation. The Mn$^{2+}$ mean spin $\langle S_z\rangle$ is directly determined from a biexciton PL spectrum detected in a circular polarization as a weighted average of the intensities of consecutive emission lines. The dependences of the above-introduced efficiency on the magnetic field (applied in Faraday configuration), and excitation power at $B=4$~T are presented in Figs. \ref{fig8}(g) and \ref{fig8}(h), respectively. Since the mean spin of the Mn$^{2+}$ ion can be independently obtained basing on the spectra detected in two circular polarizations, each of the Figs. \ref{fig8}(g), \ref{fig8}(h) contains two sets of points corresponding to $\sigma^+$ and $\sigma^-$ polarized detections. Their mutual agreement confirms the self-consistency of our experimental results.

Excitation power dependence of the ion spin orientation efficiency exhibits a pronounced decrease at high-power regime (Fig. \ref{fig8}(h)). Similarly to the case of CdTe QDs \cite{Goryca_PRL_2009}, it is related to the increased probability of the biexciton formation. More specifically, the 2X recombination leaves a randomly polarized neutral exciton in the dot. As a consequence, more frequent creation of the biexciton in the QD lowers the efficiency of the optically-induced exciton spin-transfer, which finally leads to a smaller efficiency of the ion spin orientation. On the other hand, for sufficiently low 2X formation probability assured by the excitation power corresponding to $<5\%$ of the saturation power, the spin orientation efficiency is almost power-independent. However, under these conditions it clearly increases with the magnetic field (Fig. \ref{fig8}(g)). In the low-field regime such effect can be attributed to a suppression of the mixing between Mn$^{2+}$ spin eigenstates introduced by a hyperfine interaction with nuclear spin. This mixing was previously shown to significantly increase the ion spin relaxation dynamics in the absence of magnetic field \cite{Goryca_PRL_2009_2, Goryca_PRL_2009, Kossacki_2006_Phys_E,Goryca_2006_PSSB}, which finally decreases the zero-field efficiency of optical pumping of the ion spin. However, the hyperfine coupling is relatively small \cite{Quazzaz_1995} and even a field of the order of few tenths of Tesla is sufficient to almost completely purify the ion spin eigenstates. As a result, it cannot be the underlying reason of the pronounced enhancement of spin orientation efficiency at $B>1$~T. Taking into account almost constant exciton spin-transfer efficiency in the entire range of the applied field, we interpret this enhancement as reflecting the field-dependence of the intrinsic mechanism leading to the ion spin orientation. Surprisingly, no theoretical model providing such mechanism has been proposed so far, despite the intuitive character of the optical orientation process. In particular, only the optical pumping of the ion spin under strictly resonant excitation \cite{LeGall_PRB_2010, Baudin_PRL_2011} was quantitatively understood \cite{Cywinski_PRB_2010, Cao_PRB_2011}. The deeper understanding of the ion spin orientation mediated by the injection of spin-polarized excitons would presumably also allow to explain somehow low efficiency of this process in the case of the studied CdSe/ZnSe QDs. In particular, the obtained efficiency at $B=1$~T is few times lower compared to CdTe QDs \cite{Goryca_PRL_2009}, for even higher efficiency of the exciton spin-transfer in the present case. This observation together with indication of several differences between the two types of dots, including much more efficient exciton spin relaxation for CdSe QDs, may in turn shed some light on the mechanism of the ion spin orientation and help in its further investigation.

\section{Conclusion}

We have presented a simple all-optical method enabling effective injection of spin-polarized excitons to self-assembled CdSe/ZnSe QDs. Our studies concerned both nonmagnetic dots and those containing single Mn$^{2+}$ ions. Experimentally, the pronounced exciton spin-transfer was observed under below-the-barrier excitation at 488~nm with circularly polarized light. The spin-transfer efficiency was studied in polarization-resolved measurements of a single dot PL performed in external magnetic field applied along the sample growth axis. Experimental results indicated importance of phonon-mediated neutral exciton spin relaxation occurring during the exciton lifetime in a QD. In case of nonmagnetic dots, such relaxation was simply related to X spin-flip between its Zeeman-split states leading to an increased intensity of X emission corresponding to a lower-energy state. On the other hand, the spin relaxation for Mn-doped QDs was exhibiting more complicated character. In particular, our findings revealed that it is not entirely related to X spin-flip occurring without the change of the ion spin. However, basing on the time-integrated spectroscopy we were not able to provide its detailed description.

The straightforward character of X spin relaxation in the case of nonmagnetic QD enabled us to analyze its X PL spectra obtained under circularly and linearly polarized excitation within a frame of a simple rate-equation model. On this basis we established a power-dependence of the exciton spin-flip rate on X energy splitting. Moreover, we determined the spin-polarization degree of excitons injected to QD under $\sigma^+$ and $\sigma^-$ polarized excitation to be similar in both cases and close to 40\% at magnetic field of 3-10~T. At lower field, the mixing of exciton spin states imposed by the fine structure prevented exact determination of optically-induced spin-transfer efficiency for a nonmagnetic dot. On the other hand, it was possible for Mn-doped one since in this case the fine structure is dominated by the exciton-ion exchange interaction. Consequently, we found the spin-transfer efficiency under excitation at 488~nm to be almost field-independent and equal to about 40\% in the entire range of the applied field 0-10~T . We stress out that similarly large efficiency was obtained for several, randomly selected dots exhibiting different energies of excitonic emission. This finding indicates the robustness of the presented spin-conserving excitation channel in case of the studied CdSe/ZnSe QDs.

Importantly, we observed also a negative spin-transfer for the only emission line in a CdSe/ZnSe QD PL spectrum originating from a trion recombination. Following the previous studies of different QD systems, it enabled us to unequivocally identify the negative charge state of this trion.

Finally, taking advantage of the ability to inject the spin-polarized excitons to Mn-doped CdSe/ZnSe QD we demonstrated the possibility of optical orientation of Mn$^{2+}$ spin. The ion spin readout was based on the intensities of six biexciton emission lines measured in a circular polarization of detection. Our findings revealed that under $\sigma^\pm$ polarized excitation the Mn$^{2+}$ is oriented towards $\mp5/2$ state. The efficiency of the optical orientation was found to increase with the magnetic field. Simultaneously, it was a few times lower compared to the previously studied Mn-doped CdTe QDs. The origin of both effects is not entirely clear, since the intrinsic mechanism underlying the ion spin orientation remains not fully understood. Nevertheless, the possibility of the optical control over the Mn$^{2+}$ spin embedded in a CdSe QD evidenced in our work may be further exploited to determine the ion spin relaxation dynamics in a low-field regime or even in the absence of magnetic field.

\begin{acknowledgments}
The authors thank Tomasz Kazimierczuk for stimulating discussions. This work was supported by the Polish Ministry of Science and Higher Education in years 2012-2016 as research grant ''Diamentowy Grant'', by the National Science Centre under decisions DEC-2011/01/B/ST3/02406, DEC-2011/02/A/ST3/00131, DEC-2012/07/N/ST3/03130, DEC-2013/09/B/ST3/02603, DEC-2013/09/D/ST3/03768, by the Polish National Centre for Research and Development project LIDER, and by Foundation for Polish Science (FNP) subsidy ''Mistrz''. Project was carried out with the use of CePT, CeZaMat, and NLTK infrastructures financed by the European Union - the European Regional Development Fund within the Operational Programme "Innovative economy" for 2007 - 2013. 
\end{acknowledgments}

\end{document}